\documentclass[preprint2]{aastex}

\shorttitle{Emission Flickering from the Secondary Star in CVs? V3885 Sgr}
\shortauthors{Ribeiro, F.M.A, Diaz, M.P.}

\begin{document}

\title{Emission Line Flickering from the Secondary Star in Cataclysmic Variables? A study of V3885 Sagitarii}

\author{Fab\'{i}ola M. A. Ribeiro and Marcos P. Diaz}
\affil{Instituto de Astronomia, Geof\'{i}sica e Ci\^{e}ncias Atmosf\'{e}ricas, Universidade de S\~{a}o Paulo,\\ r. do Mat\~{a}o 1226, cid. Universit\'{a}ria, cep 05508-900, S\~{a}o Paulo, SP, Brazil}
\email{fabiola@astro.iag.usp.br, marcos@astro.iag.usp.br}

\begin{abstract}
Spectrophotometric observations of H$\alpha$ and He\begin{small} I\end{small} 6678 emission lines of the nova-like Cataclysmic Variable V3885 Sgr are presented and analyzed. The binary orbital period was determined as P = 0.20716071(22) days. Doppler Tomography was performed with both H$\alpha$ and He\begin{small} I\end{small} lines. Disc radial emissivity profiles were also computed. The tomography mapping of flickering sources was performed using the H$\alpha$ line, from which we concluded that the flickering is not uniformly distributed on the disc. The observed tomogram of the flickering was compared with simulations, suggesting that the most intense flickering source in the H$\alpha$ is not located in the accretion disc. It is proposed that the main line flickering source may be associated with the illuminated secondary star.
\end{abstract}

\keywords{accretion, accretion disks --- binaries: close --- cataclysmic variables --- stars: individual (\objectname{V3885 Sgr})}

\section{Introduction}

Cataclysmic Variables (CVs) are close binary systems composed of a white dwarf, also called primary, and a red dwarf star, the secondary. The secondary fills its Roche lobe and transfers matter to the primary. Due to the binary angular momentum, the matter does not fall toward the white dwarf, forming an accretion disc. The accretion disc emission is complex, it has regions of enhanced emission like the hot spot, the locus where the stream of matter coming from the secondary reaches the disc, as well as the boundary layer, the inner part of the accretion disc where the matter is decelerated to the white dwarf equatorial velocity. The disc's vertical profile is not flat; it presents a higher thickness in its outer parts. A review about CVs is presented by \citet{War95}.

The CVs are grouped into many classes and subclasses; one of these is the nova-like class that harbor non-eruptive objects. V3885 Sgr is one of the brightest CVs. It is a non eclipsing system classified as an UX UMa nova-like star. V3885 Sgr was discovered by \citet{Bid68} who detected broad Balmer emission lines. \citet{Bon71} identified the continuum slope as similar to those of DA white dwarfs. \citet{Weg72} observed emission lines superposed on absorption lines identifying the object as a nova-like CV. \citet{Hes72} first noticed the occurrence of rapid photometric variability in the system. \citet{Gui82} carried a spectroscopic study in the UV noticing the existence of P-Cygni profile in the C\begin{small} IV\end{small} line, which indicates the presence of a wind in the system. From UV, optical and IR continuum distributions, \citet{Hau87} described the spectral energy distribution of the disc as one of an stationary disc with high mass transfer rate. \citet{Har05} calculated P$_{orb}$ = 0.207135(15) d for this system, solving previous inconsistencies concerning the orbital period. They also presented a radial velocity study, H$\beta$, H$\gamma$ and HeI 4471 Doppler maps, and they showed evidences of irradiation for the secondary star.

Flickering or rapid variability is observed as a stochastic variation of the emitted radiation, with time-scales from seconds to tens of minutes and amplitude ranging from a few percent of magnitude to more than one magnitude. Flickering is associated with mass transfer and occurs in objects other than CVs, like symbiotic stars \citep{Mik90} and X-ray binaries \citep{Bap02}. Furthermore, flickering does not only occur in systems where an accretion disc is present but it is also observed in magnetic CVs. The flickering in the continuum could be produced in different regions of the system. \citet{War71} observed that the flickering in U Gem disappeared during the eclipse, and since the central part of the disc is not occulted during the eclipse, they associated the flickering source with the hot spot. \citet{Vog81} verified that the flickering in OY Car persists during the hot spot eclipse. The same was observed for HT Cas by \citet{Pat81} and Z Cha by \citet{Woo86}. \citet{Hor85} identified the flickering in some eclipsing systems as being originated near the center of the accretion disc. In addition, \citet{Hor94} associated the flares observed in OY Car with the innermost part of the disc. \citet{Bru96} observed that the flickering forming region in Z Cha is located near the white dwarf and that there could be other flickering sources seen in different photometric states.

In the next section we detail the observations and data reduction procedure. In section 3 the average spectrum is described and a long term ephemeris is given. The system masses and orbital inclination are constrained. Doppler Tomography and Flickering Doppler Tomography results are presented. Furthermore, the observed flickering tomograms are compared with flickering simulations. The implications of our results are discussed in section 4. The conclusions are outlined in section 5.

\section{Observations and Data Reduction}

The time-resolved spectroscopic observations were made using the 1.60-m telescope at Observat\'orio Pico dos Dias (LNA/CNPq) and the 1.5-m telescope at Cerro Tololo Inter-American Observatory. For more details about the observations, see the observations journal (Table 1)\notetoeditor{Table 1 must be a double column table}. The atmospheric conditions on the observing runs ranged from photometric to spectroscopic. Differential spectrophotometric techniques were employed. The LNA observations were made using a Cassegrain spectrograph, equipped with a 1200 grooves/mm grating blazed at 4550 \AA{}, and a back-illuminated 1024x1024 CCD. The CTIO observations were also carried out with a Cassegrain spectrograph, a 1200 grooves/mm grating, but blazed at 7500 \AA{}, and a 1200x800 Loral 1K CCD. All the observations were centred at the H$\alpha$ line. A spectral resolution of about 2 \AA{} was achieved. The spectral coverage also included the He\begin{small} I\end{small} 6678 spectral line.

\placetable{tbl-1}

Bias and flatfield images were taken to adjust the zero level count and to correct for response. When the observational conditions allowed, twilight flat images were also taken for deriving the CCD illumination correction. The data were bias subtracted, flatfielded and illumination corrected using standard IRAF\footnote{IRAF is distributed by the National Optical Astronomy Observatories, which are operated by the Association of Universities for Research in Astronomy, Inc., under cooperative agreement with the National Science Foundation} procedures. The spectra were extracted using an optimal extraction algorithm \citep{Hor86}, for both the object and a comparison star. Each spectrum was then corrected for possible cosmic rays by interpolating neighbor pixels. We set the optimal extraction to trim only the most intense cosmic rays, as a sharp line profile feature (i.e. due to flickering) could be confused with a cosmic ray and trimmed. So we prefer to eliminate the less intense cosmic rays by hand. The cosmic rays in the sky background regions were all automatically removed. The wavelength calibration was performed by intercalating the Ne-Ar arc lamp spectra with observations of the variable star, resulting in an accuracy of $\sim$ 0.05 \AA{} RMS or 2 km s$^{-1}$. The interval between arc-lamp exposures was chosen to minimize the spectrograph mechanic flexure effects over the wavelength scale. Wide slit tertiary standard photometric stars \citep{Ham94} observations were taken for flux calibration. To allow the correction of light losses at the spectrograph's slit and to perform small sky transparency corrections, a nearby comparison star was included in the slit. Wide slit (7 arcsec) spectra of the comparison star were also acquired. This differential calibration procedure also corrects the spectra from atmospheric dispersion effects.

As the H$\alpha$ line region includes telluric lines and bands, we have performed a correction using a telluric template with the same spectral resolution. The telluric template was scaled and shifted to correct the telluric features using an averaged spectrum of the target. This correction was then applied to each spectrum of our dataset. For this reason, the correction should be good enough to the average spectrum and may not be so efficient for each individual spectrum. Individual correction factors for of each spectra are unreliable because of their low S/N ratio. All derived quantities involve averaging data on phase bins therefore we do not foresee any impact on the measured quantities.

\section{Results}

\subsection{Spectral Features}

The averaged spectrum (fig. 1)\notetoeditor{figure 1 must be a single column figure} has a spectral coverage of 6400-6714 \AA, showing H$\alpha$ and He\begin{small} I\end{small} 6678 emission lines. The continuum has a blue slope, without any other visible feature. H$\alpha$ appears as single-peaked and He\begin{small} I\end{small} as double-peaked. The former line presents a small absorption component at the blue side of the profile, indicating a P-Cygni profile or an underlying absorption. The H$\alpha$ line has a FWHM of 780 km s$^{-1}$ , $EW = -4.1$ \AA{}, and the line profile integrated flux $F$ is $5.7~10^{-13}$ erg s$^{-1}$ cm$^{-2}$ while the He\begin{small} I\end{small} 6678 line has a FWHM of 820 km s$^{-1}$, $EW = -0.48$ \AA{} and  integrated flux $F = 6.40 10^{-14}$ erg s$^{-1}$ cm$^{-2}$ thus the equivalent width of H$\alpha$ line is 8.5 times the He\begin{small} I\end{small} one.

\begin{figure}
 \includegraphics[width=75mm,angle=270]{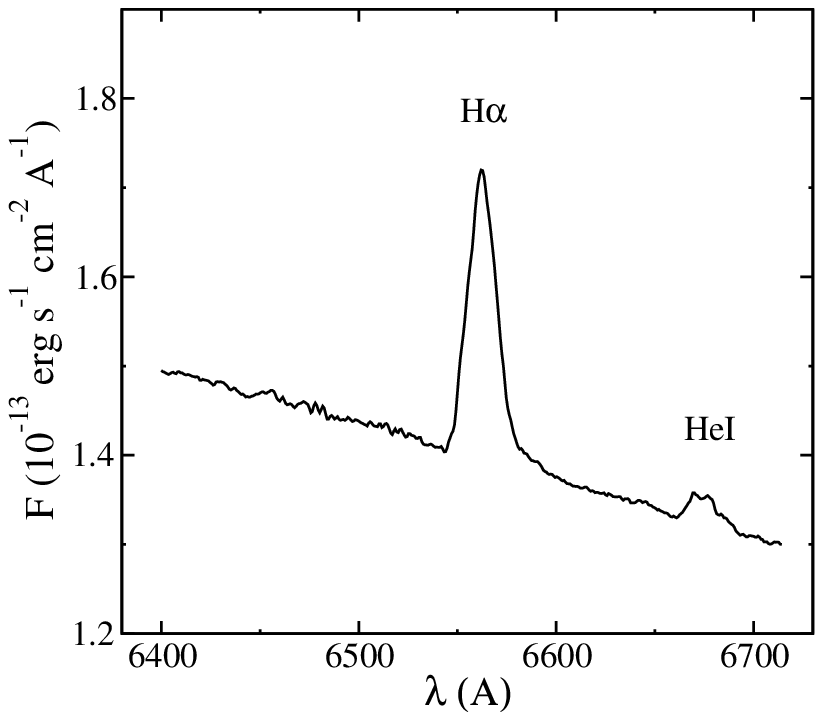}
 \caption{V3885 averaged spectra. The H$\alpha$ and HeI lines are indicated.}
\label{fig1}
\end{figure}

\placefigure{fig1}

It was not possible to identify any absorption spectral feature from the secondary in the phased spectra or in the averaged spectra. So, the secondary radial velocity amplitude $K_2$ and the companion spectral type could not be directly determined in the present work.

\subsection{System Parameters}

A long term ephemeris was derived using unambiguous cycle-counting between all observation nights in our dataset. The calculated orbital period for V3885 Sgr is P$_{orb}$ = 0.20716071(22) d and epoch E$_0$ = 51942.08371(46) MHJD, defined as the positive to negative crossing of radial velocity curves for the line wings. The uncertainties were obtained from a least squares linear fit to all cycle timings. The residuals in the O-C diagram are well behaved and there is no suggestion of a period derivative. This period is in agreement with the 0.207135(15) d period found by \citet{Har05} within a $2\sigma$ interval.

Using the \citet{Sch80} method to measure the velocity in the line wings, we have constructed a diagnostic diagram for both H$\alpha$ and HeI as a function of the convolution mask Gaussians separation $a$. The primary radial velocity ($K_1$) values remained almost constant inside the interval $300 < a < 600$ km s$^{-1}$, presenting a slightly decrease in velocity for $a > 600$ km s$^{-1}$. The relative error on $K_1$ as a function of $a$ showed a minimum at $a \approx 500$ km s$^{-1}$. The RMS exhibited this same behavior. From this diagnostic diagram, looking for a range in $a$ that presents a plateau in $K$ and minimum relative error and RMS, we estimated the primary radial velocity from  $K_{H\alpha}$ = 139(25) km s$^{-1}$ for the H$\alpha$ emission line and $K_{HeI}$ = 188(25) km s$^{-1}$ for He\begin{small}I\end{small}. These uncertainties were estimated on the basis of the assumption that the radial velocity measurements from both H$\alpha$ and He\begin{small}I\end{small} should be compatible. The $K_1$ value does not vary significantly with the Gaussian mask width. The Gaussian masks FWHM widths were 200 km s$^{-1}$ for H$\alpha$ and 150 km s$^{-1}$ for He\begin{small}I\end{small}. These radial velocity values are compatible with $K$ = 166(6) km s$^{-1}$ presented by \citet{Har05}. The systemic velocity was also derived for both lines, $\gamma_{H\alpha}$ = -43(25) km s$^{-1}$ and $\gamma_{HeI}$ = -84(25) km s$^{-1}$.

The double-Gaussian convolution technique for radial velocity measurements is not immune to systematics errors \citep{Und06}. In some cases it produces wrong masses estimates, even with large datasets like the one present here. Unfortunately, there is no other way to constrain the primary mass in this non-eclipsing nova-like CV.

\begin{figure}
 \includegraphics[width=80mm]{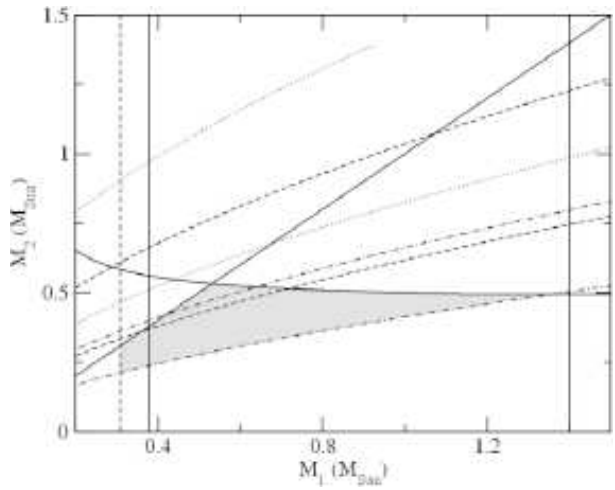}
 \caption{Mass diagram. For a description about the lines see the text. The region of most probable masses is shown in gray.}
 \label{fig2}
\end{figure}

\placefigure{fig2}

The mass diagram for V3885 Sgr (fig. 2)\notetoeditor{figure 2 must be placed as a single column figure} was constructed using the H$\alpha$ radial velocity semiamplitude and its error. The continuous vertical line at the right indicates the white dwarf upper mass limit. The two vertical lines at the left are the white dwarf lower mass limit for i = 75$\degr$ (dashed line) and i = 45$\degr$ (continuous line). These limits were obtained from each line FWZI, considering a Keplerian disk and a mass-radius relationship \citep{Had61} for the primary star. The diagonal straight line is the equal mass line, plotted only as a reference as the stable mass transfer line is not straight along the mass diagram and it is also model dependent. The quasi-horizontal continuous line represents the upper mass limit for a main sequence secondary filling its Roche lobe. The $M_1$,$M_2$ relation is plotted for a few inclination values (i = 35$\degr$ is plotted as the dotted lines, i = 45$\degr$ as the dashed lines and i = 75$\degr$ as the dot-dashed lines), considering $K_1$ and its uncertainty. The region of most probable masses is shown in gray. The primary mass is unconstrained between 0.3 M$_{\sun}$ and approximately 1.4 M$_{\sun}$, and the secondary mass is possibly between 0.2 and 0.55 M$_{\sun}$. The primary mass range obtained by \citet{Har05} is $0.55 < M_1 < 0.8$ M$_{\sun}$. Another parameter that can be constrained from the mass diagram is the orbital inclination, which is found between 45$\degr$ and 75$\degr$. The analysis and conclusions in the following sections are not dependent on a precise $M_1$ value.

\subsection{Doppler Tomography}

The Doppler Tomography method, proposed by \citet{Mar88}, uses the velocity information from spectral line profiles at various orbital phases in order to reconstruct the binary system's emissivity in the velocity space. The method is based on a few hypothesis listed below. All the emission line production regions must be always visible and their line optical depth must be negligible, which is violated if observations during eclipses are used. The line flux at any point of the binary rotating frame of reference must be constant. Outbursts represent a flux variation, breaking this hypothesis. All the motions must be parallel to the orbital plane. Wind originating from the disc has a velocity vector out of the orbital plane. All the velocity vectors rotate with the binary system. The last hypothesis is that the intrinsic line profile width in any point is negligible; saturation effects at large optical depths can widen the line profiles \citep{Hor95}.

The Doppler tomograms from the H$\alpha$ and He\begin{small} I\end{small} line profiles were  obtained using the filtered back-projection method \citep{Ros82} and are shown in Fig. 3\notetoeditor{figure 3 must be a double column figure}. The coordinate system is defined with the $x$-axis pointing from the primary to the secondary and the $y$-axis pointing in the direction of the secondary's motion. Considering a Keplerian disc, the outer part of the tomogram (higher velocities) represents the inner region in the position space, and vice-versa. The H$\alpha$ and He\begin{small} I\end{small} tomograms were calculated using 1888 independent spectra. A 70 km s$^{-1}$ FWHM Gaussian convolution filter was applied to the H$\alpha$ line and a 120 km s$^{-1}$ one was applied to the He\begin{small} I\end{small} line. The tomogram resolution is a combination of the spectral resolution, convolution filter width and phase sampling.
As the V3885 Sgr masses and orbital inclination cannot be strictly determined, the marked positions on the tomograms are plotted just as a reference. Distinct values of orbital inclination and masses would produce a rescaling of the marked positions.

\begin{figure}
\includegraphics[width=160mm]{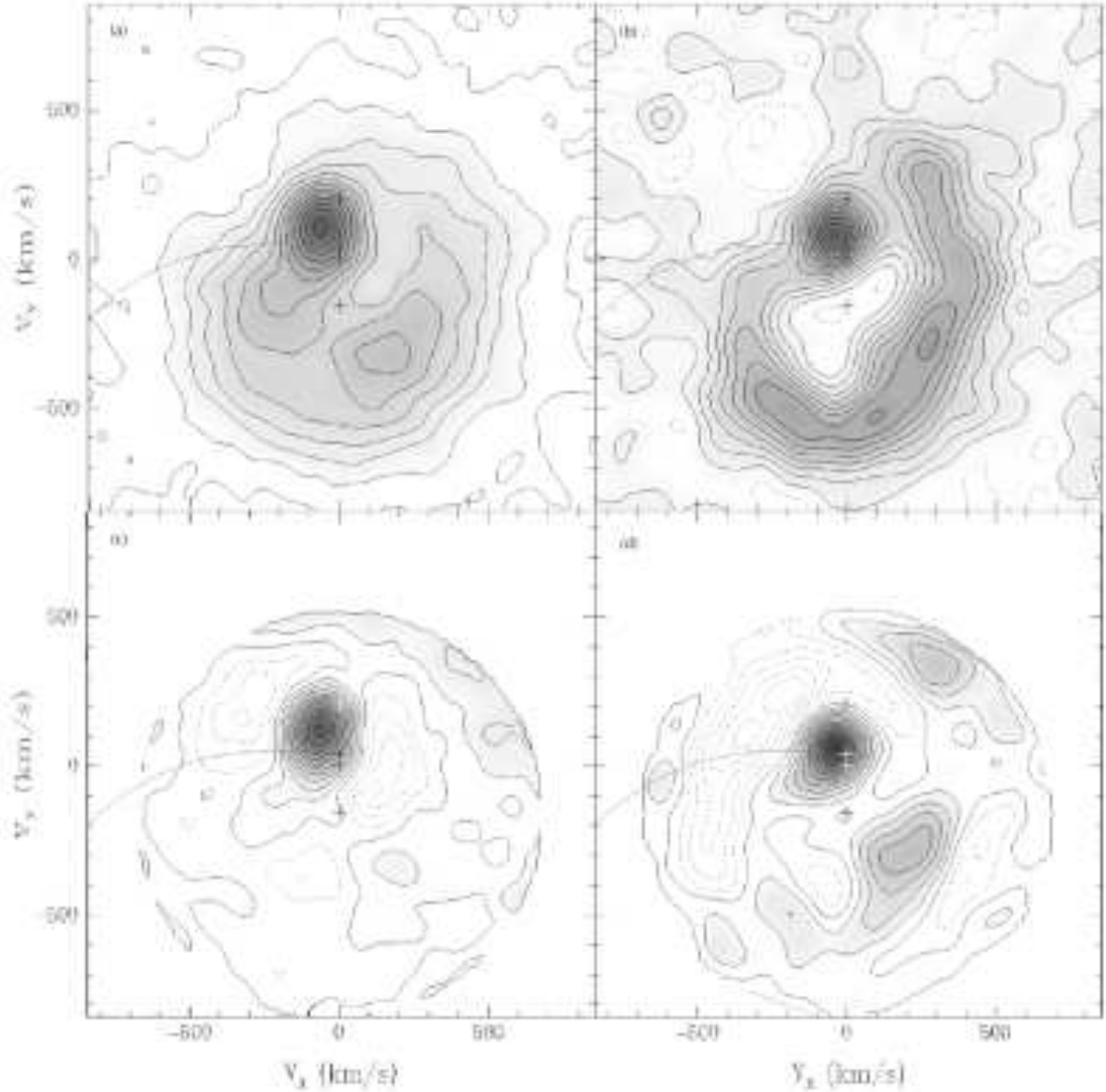}
\figcaption{Top: (a) H$\alpha$ and (b) He {\scriptsize I} Doppler Tomograms. The central '+' marks the system's center of mass, above it the internal Lagrange's point $L_1$ and the secondary's center of mass are marked. Below the origin is the center of mass of the primary.
These positions are calculated for M$_1 = 0.57$ M$_\odot$, M$_2 = 0.42$ M$_\odot$ and 60$\degr$ orbital inclination. The estimated resolution for the tomograms are 111 $km~s^{-1}$ (FWHM) for H$\alpha$ and 147 $km~s^{-1}$ (FWHM) for He {\scriptsize I}. Bottom: Tomograms of (c) H$\alpha$ and (d) He {\scriptsize I} emission centred at the primary and subtracted by a radially symmetric disc fitting.\label{fig3}}
\end{figure}

\placefigure{fig3}

In the H$\alpha$ tomogram (Fig. 3a) one can see that the most intense emission is located at the $(-V_x, +V_y)$ quadrant, and we could not affirm based only on this map whether this emission is originated from the illuminated secondary or from the hot spot. It is not aligned with the position of the stellar centres or clearly over the stream trajectory. The stellar positions could move along the y axis direction depending on the actual orbital inclination. An extended region in the $(-V_x,-V_y)$ and $(+V_x,-V_y)$ quadrants can also be seen, corresponding to disc regions opposite to the secondary star. In order to show more clearly these tomogram anisotropies, a radially symmetric disc image was subtracted from the maps (Fig. 3).

The He\begin{small} I\end{small} tomogram (Fig. 3b) presents a ring shape when compared with H$\alpha$, suggesting that the emission from the former is mostly produced in the disc itself. The spot in the $(-V_x, +V_y)$ quadrant on the HeI map is approximately at the same location as the H$\alpha$ map spot. The asymmetric component is more structured in He\begin{small} I\end{small} than in H$\alpha$, and clearly shows the enhanced emission in the $(+V_x,-V_y)$ quadrant.

Comparing the H$\alpha$ and He\begin{small} I\end{small} tomograms (Fig. 3), one can see that the He\begin{small} I\end{small} emission is more intense in the $(+V_x,+V_y)$ quadrant and in the lower part of the $(-V_x,-V_y)$ quadrant. The origin of such behavior is still unclear. It can also be seen that the H$\alpha$ emission is more intense than the He\begin{small} I\end{small} one in the $(-V_x, +V_y)$ quadrant. The emission from the $(+V_x, -V_y)$ quadrant in both lines can be explained if the stream does not hit the disc in the $(-V_x, +V_y)$ quadrant, but it flies above the disc, reaching the disc in the $(+V_x, -V_y)$ quadrant, before the point of minimum approach to the primary. Therefore, part of the stream material is not incorporated to the disc in the hot spot region, but follows a free particle trajectory above the disc until it reaches the $(+V_x, -V_y)$ quadrant. This hypothesis is plausible if one reminds that the accretion disc geometry is not flat, and so the height in the disc's outer rim is greater than in the central regions and, in addition, the primary gravity has an increasing vertical component for smaller radii.

The consistency of the Doppler maps was verified by comparing trailed spectra generated from the projection of the maps at many orbital phases and the trailed spectra generated from the observed line profiles (Fig. 4)\notetoeditor{figure 4 must be placed as a double column figure}. The comparison between the phase maps suggests a reasonable consistence of the tomographic reconstructions. A significant difference between the maps is related to the sharp and low velocity feature seen in the observed phase
maps. This feature appears once per orbit in the observations while it is spread into two features on each orbit in the projected map. As the tomography algorithm considers that all the structures are visible all the time, this sharp emission should be related to the illuminated face of the secondary star (see discussion in section 3.5).

\begin{figure}
\includegraphics[]{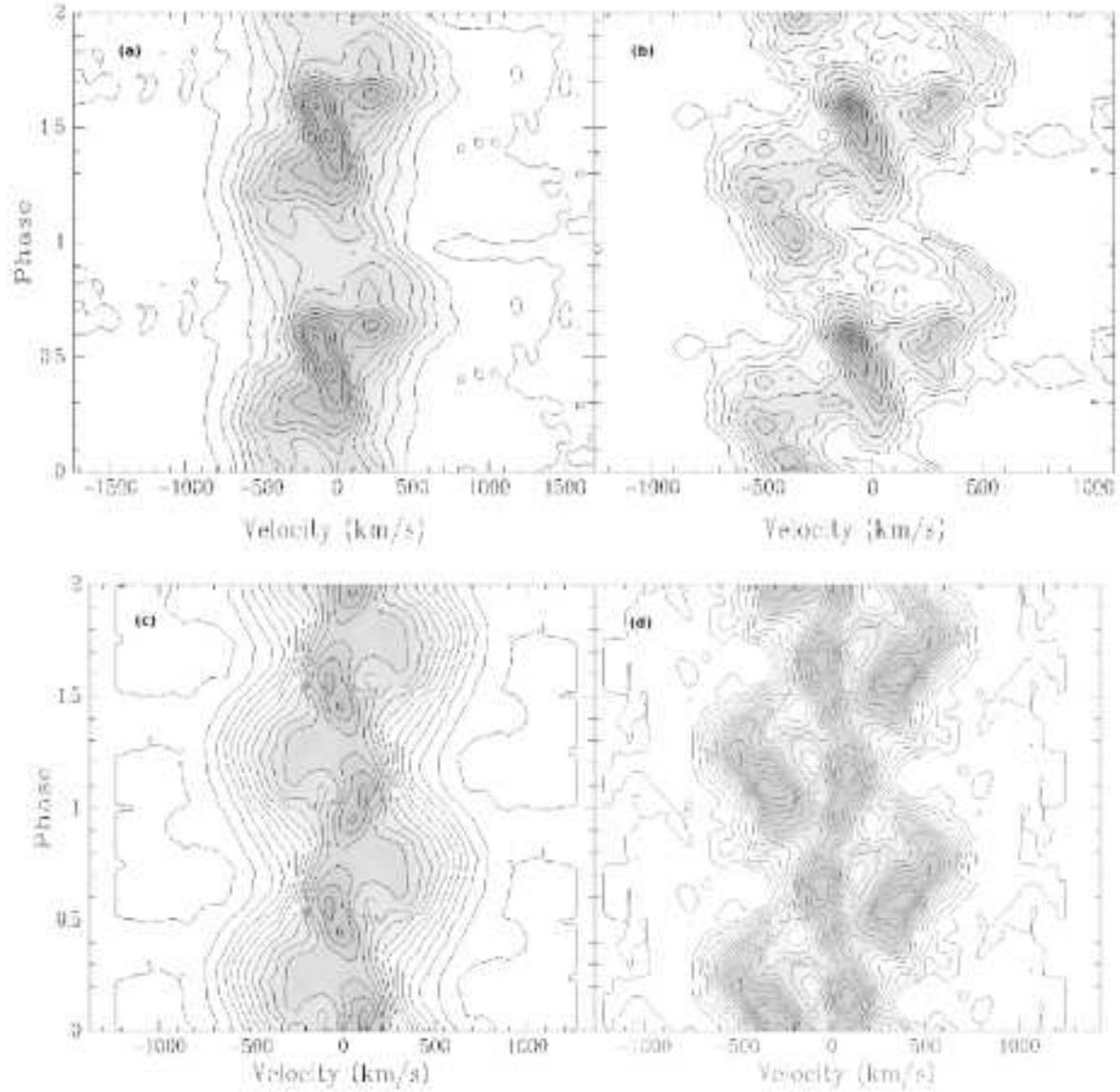}
\figcaption{Comparison between phase maps constructed from observed (a) H$\alpha$ and (b) HeI 6678 line profiles and (c) H$\alpha$ and (d) HeI phase maps from line profiles reconstructed from the Doppler maps. Notice that the scales of the velocity axes are distinct for each graph panel.\label{fig4}}
\end{figure}

\placefigure{fig4}

Emissivity profiles in velocity space can be obtained directly from the Doppler tomograms. That was done using Doppler tomograms for H$\alpha$ and He\begin{small} I\end{small} centred at the primary's position (Fig. 2b)\notetoeditor{figure 5 must be a double column figure}. The emissivity profiles are similar to line profiles but they are free from the projection of intrinsically high velocity emission that may appear at lower velocity in the line profiles. In order to produce the velocity profiles, the intensity along each concentric ring centred at the primary were represented by its mode. With this procedure, we verified that the H$\alpha$ tomogram is more homogeneous than the He\begin{small} I\end{small} one, with the mode, median and mean yielding similar results.

\begin{figure}
\includegraphics[angle=270, width=150mm]{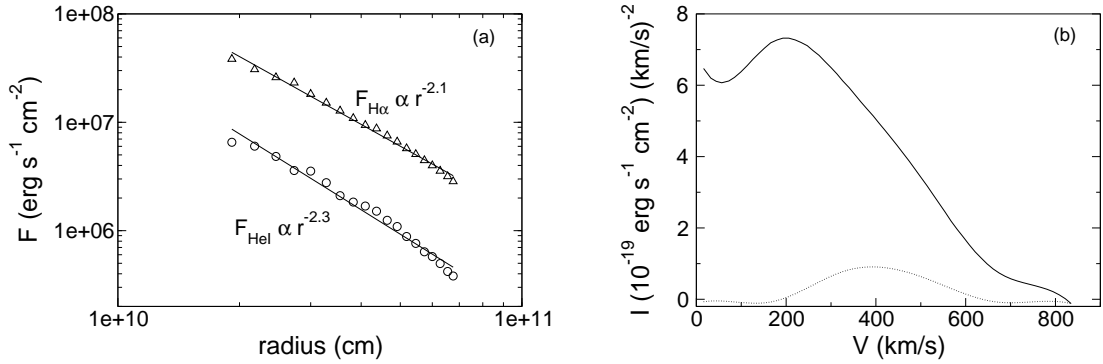}
\figcaption{(a) Radial emissivity profile for H$\alpha$ (triangles) and He {\scriptsize I} (circles) corrected from reddening. Simple power laws are fitted to the data. (b) Emissivity profiles as a function of the velocity of the emitting region, for both H$\alpha$ (solid line) and He {\scriptsize I} (dotted).\label{fig5}}
\end{figure}

\placefigure{fig5}

Doppler tomograms are always plotted in the velocity space, but one can transform the maps to position space, assuming a Keplerian velocity regime. One problem that could occur in this transformation arises from the fact that the disc's most internal regions (seen in the line far wings) are brighter and carry the larger uncertainties in the inversion process. To study the disc emissivity profiles, the emission at a given radius was estimated, obtaining the disc radial emissivity profiles for both H$\alpha$ and He\begin{small} I\end{small} (Fig. 3a). This value is calculated as the mode along each ring because it is less susceptible to local variations due to the disc emission anisotropies. To obtain the absolute radial emissivity profiles, we used a distance to the system equal to 110(23) pc from \textit{Hipparchos} parallax \citep{ESA97} and a color excess of $E(B-V)$ of 0.02 \citep{Bru94}. The profiles are shown in Fig. 5. The radial emissivity profiles may be fitted by power laws, $I_{H\alpha}~\alpha~r^{-2.1}$ and $I_{He I}~\alpha~r^{-2.3}$. In this system the He\begin{small} I\end{small} emission seems more concentrated in the inner disc regions than the H$\alpha$ one. The obtained radial emissivity indexes are similar to those obtained for V841 Oph by \citet{Dia03}. Radial emissivity profiles were also obtained from Doppler maps for CU Vel \citep{Men96}, V347 Pup \citep{Dia99}, RR Pic \citep{Dia03} and U Gem during outburst \citep{Mar90}.

\subsection{Flickering Tomography}

We have explored until this point the "stationary" line emission from V3885 Sgr. The line emission intrinsic variability can be conveniently studied using the tomographic reconstruction of the flickering emission. In the flickering tomography method \citep{Dia01}, we start calculating the variance of the observed line profiles into phase bins. Then, the intrinsic line variability is isolated from the instrumental, orbital, and secular components. The instrumental component of the variance is calculated as a combination of the Poisson noise and a Gaussian readout noise. The orbital component contribution is considered negligible, since the data is sampled into small phase bins. The secular component is the remaining long term variability of the system. The effect of such a component was found to be negligible by replacing the long term average profiles by profiles computed from individual runs. Doppler tomograms are calculated from these intrinsic variance line profiles mapping the line emission variability in the system. The flickering tomography method is very demanding on the amount of collected data and also sensitive to the time sampling.

Pure photon and readout noise were removed from the data using the empirical noise model described above. However, residual noise from other sources (such as continuum subtraction and differential calibration errors) are still present in the variance data.

A total of 1863 spectra from our original data set were used in the flickering tomography study. Some spectra were not used in this analysis because they deteriorate the statistics on the quantities calculated with these data. The flickering tomography could only be performed with the H$\alpha$ line, which is more intense than the He\begin{small} I\end{small} one. In Fig. 6\notetoeditor{figure 6 must be a single column figure} we show the average of all spectra used in flickering tomography and the average RMS spectrum already corrected from instrumental, orbital as well as secular effects. The flickering line profile defined as above is single peaked. The RMS increase with the emission intensity but not as expected from pure photon noise. It can be seen that the flickering and the flux line profiles present slightly distinct intensities in the blue side continuum region. The RMS flickering intensity relative to the average line intensity obtained in this work is larger than that observed in the nova-like V442 Oph \citep{Dia01}.

\begin{figure}
\includegraphics{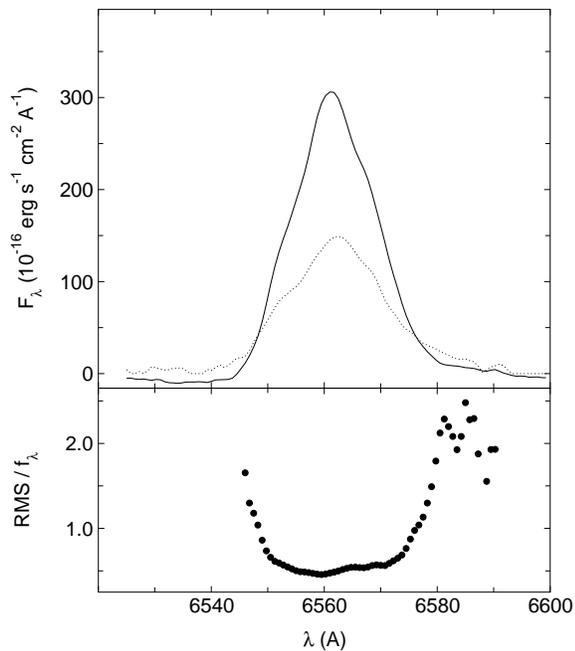}
\figcaption{At the top: averaged flux (solid line) and flickering RMS (dotted line) spectra. In the lower panel the ratio between these profiles is presented.\label{fig6}}
\end{figure}

\placefigure{fig6}

The H$\alpha$ flickering Doppler tomogram is shown in Fig. 7\notetoeditor{figure 7  must be a single column figure}. The filter used to produce this map was set to 100 km s$^{-1}$ FWHM and the final tomogram resolution is estimated as 130 km s$^{-1}$. The positions marked in the tomogram are, as in the flux Doppler tomograms, from top to bottom, the secondary center of mass, the internal Lagrange's point $L_1$, the system's center of mass, and the center of mass of the primary. As the orbital inclination and masses cannot be strictly determined, the stellar positions are plotted just as a reference.

\begin{figure}
\includegraphics[width=80mm]{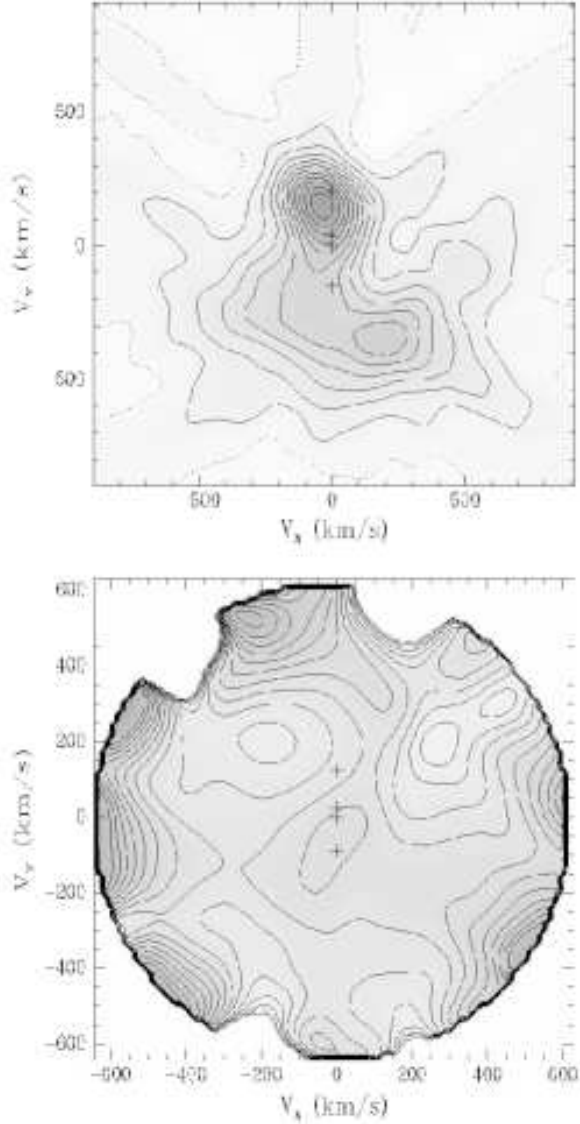}
\figcaption{(a) Flickering variance tomogram and (b) the ratio between the flickering and flux tomograms. The '+' are, from top to bottom, the center of mass of the secondary, the $L_1$ Lagrange's point, the system's center of mass and the center of mass of the primary. These positions are calculated for M$_1 = 0.57$ M$_\odot$, M$_2 = 0.42$ M$_\odot$ and 60$\degr$ orbital inclination. The high amplitude features in the border of the lower panel map are artifacts produced by the division by values close to zero (the available signal in both flux and flickering tomograms drops with increasing velocity). The velocity of the ballistic stream at the tidal disk radius is marked with a 'x'.\label{fig7}}
\end{figure}

\placefigure{fig7}

At least two regions of enhanced variance can be identified: (i) the $(-V_x,+V_y)$ quadrant near the line that connect the stellar centres, and (ii) the $(+V_x,-V_y)$ quadrant. Comparing this variance tomogram with the flux tomogram (Fig. 3), one can see that the region of higher emissivity has a different shape and it is significantly shifted in velocity.

The ratio of the RMS flickering tomogram and the flux tomogram was calculated (Fig. 5) in order to verify if the flickering is dependent of the line flux of the originating region. The high velocity regions of the resulting image were also excluded because the flux tomograms are more noisy and present values near zero. The resulting relative tomogram is then limited to intrinsic velocities below $\approx 400$ km s$^{-1}$. The ratio between the tomograms does not present bright regions as do the flux and variance tomograms. However, the relative flickering map is not flat but regions of lower intensity and a higher intensity near the center of the image can be seen. From this it can be concluded that the flickering can vary with the region of the disc where the flickering is formed, and that the flickering RMS activity may be independent of the flux intensity of its originating region. Additionally, the flickering in V3885 Sgr is not uniformly distributed in the disc. This behavior was also observed in V442 Oph by \citet{Dia01}.

\begin{figure}
\includegraphics[width=150mm]{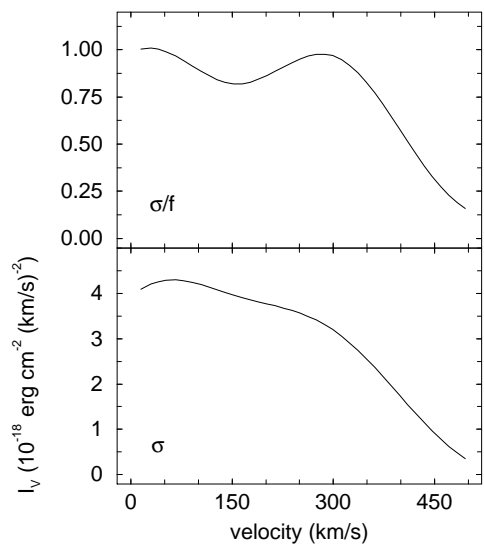}
\figcaption{Flickering activity profiles as a function of the their formation velocity. The RMS curve is given in the top panel, while the ratio between RMS and flux is shown in the bottom panel.\label{fig8}}
\end{figure}

\placefigure{fig8}

In order to obtain flickering activity profiles, variance tomograms centred in the primary were also calculated, excluding the high noise peripheral regions. Following the same method used for the calculation of the emission profiles, the median along concentric rings centered on the primary was calculated. These emissivity profiles as a function of velocity are shown in Fig. 8\notetoeditor{figure 8 must be a single column figure}. The median was used as the statistical criterion because it showed to be more robust than the average or mode, avoiding that the emissivity profile could be affected by the high noise regions. One can see from the upper panel in Fig. 8 that the averaged flickering at a given velocity (or radius in a Keplerian disc) is roughly correlated with the disc emissivity for absolute velocities higher than approximately 300 km s$^{-1}$.  These profiles have different behavior when compared with those found for V442 Oph, where both the emissivity profiles and the ratio of the tomograms presented a depression at lower velocities. In V3885 Sgr such behavior is not observed.

\subsection{Flickering Simulations}

Simulations of accretion discs including a stochastic variability component were made, aiming to perform a better analysis of the flickering tomogram. The simulations were held considering a Keplerian accretion disc, with a line emissivity profile given by a radial power-law. For each `exposure', a synthetic line profile was generated, considering the presence of noise with a given S/N ratio. Over each synthetic line profile, the contributions of isolated emission sources in the disc can be included to simulate the hot spot and/or the boundary layer. The synthetic flux and flickering tomograms were generated with the resulting line profiles.

The flickering activity was modeled as a series of flares at random positions inside a restricted region of the disc, or at the secondary's illuminated face. The flickering mean frequency and amplitude are given as a function of the total line profile flux. These are the main simulation parameters. The flickering flares were included in the simulation as having instantaneous rise and exponential decay.

Aiming to verify whether the flickering is produced in the accretion disc, a tomogram of a Keplerian disc was simulated using the V3885 Sgr orbital period and probable masses. Comparing this tomogram with the observed flickering tomogram, one can verify that the position of the most intense flickering feature (hereafter C$_1$) does not match the accretion disc in the tomogram (Fig. 9a)\notetoeditor{figure 9 must be a double column figure, it is also a color figure}. The simulations show that the absolute velocity of the flickering $C_1$ source is not consistent with the velocities found in any Keplerian disk contained within the primary Roche lobe, moreover a tidal limited disk. However, the possibility of the $C_1$ feature being originated in a non-Keplerian disk cannot be excluded at all. The absolute ballistic velocities at hot spot impact region ($v_{stream} \approx 280$ km s$^{-1}$ at the tidal disk radius - see fig. 7) are large when compared with the $C_1$ absolute velocity ($\approx 130$ km s$^{-1}$), as expected from the gas accelerating in the white dwarf's potential well.If the actual disk radius is smaller than the tidal radius
such a difference would be even larger. These two arguments suggest that $C_1$ is not the result of line flickering at the hot spot or outer disk and they are independent of the addopted orbital phase.

The next step is to test the hypothesis of the C$_1$ feature originated in the secondary's illuminated face. Synthetic flickering tomograms were computed with flickering flares from the secondary's inner face and these maps were compared with the observed flickering tomograms (Fig. 9b). We can verify a fairly good agreement between these two reconstructions. This simulation was repeated considering different values of stellar masses and orbital inclination given by the mass diagram, and the position agreement between the flickering feature $C_1$ and the inner face of the secondary star in the tomogram remained.

\begin{figure}
\includegraphics[]{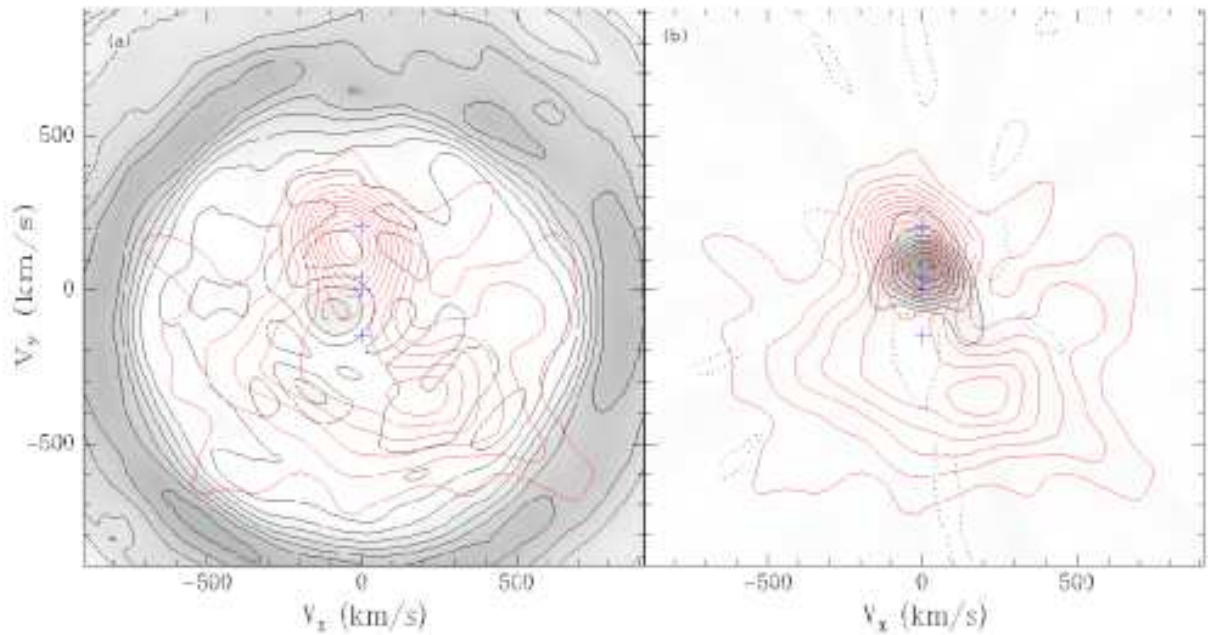}
\figcaption{(a) Synthetic accretion disk map and (b) synthetic tomogram of flickering coming from the secondary (greyscale and black contour levels) superposed with the observed flickering tomogram (red contour levels). The stellar positions shown in blue are the same of figure 6.\label{fig9}}
\end{figure}

\placefigure{fig9}

There are at least two possibilities to explain the observation of line flickering at the companion star: first, the variability could be due to the illumination effect by flickering events produced at the accretion disc or, second, they could be due to fast chromospheric activity of the secondary. Below we will examine these two possibilities.

To verify the plausibility of the C$_1$ flickering feature being observed in the secondary due to the reprocessing of the flickering from the disc, some basic quantities are calculated. The typical recombination time-scale is given by equation 1 below, where $N_e$ is the electronic density and $\alpha_A$ is the recombination coefficient. An electron density $N_e \approx$ 3x10$^{12}$ cm$^{-3}$ is need in order to observe flickering with a time-scale of about 1 second.

\begin{equation}
{\tau_r} \approx {1 \over {N_e \alpha_A}} \approx {{3\times10^{12}} \over {N_e}}  s \approx {{10^5} \over {N_e}}  yr
\end{equation}

The recombination volume needed to produce a given flux $F$ by recombination is given by equation 2, where $d$ is the distance to the source, $\nu$ the frequency of the emission and $\alpha$ the recombination coefficient.

\begin{equation}
V = {{16 \pi^2 d^2 F} \over {\nu \alpha N_e^2}}
\end{equation}

The flux of the C$_1$ feature in the observed flickering tomogram was obtained by integrating the emission in the feature and subtracting the flux of the diffuse emission near it, yielding $F_{C_1} = 3.6 \times 10^{-14}$ erg s$^{-1}$ cm$^{-2}$. In addition, a constraint on the electron density is needed. A lower limit could be obtained from the fact that forbidden lines are not observed in CVs, from this, the electronic density must be higher than $\approx$ 10$^9$ cm$^{-3}$.

Using equation 2, the lower electronic density limit mentioned above gives a volume of $V \approx 10^{34}$ cm$^3$, that is even bigger than the secondary's volume \citep{Egl83} so the recombination process could not generate the given flux in an environment with $N_e$ = $10^9$ cm$^{-3}$.

An upper limit to the average density can be obtained from the classical atmosphere models by \citet{Hub95}. The structure of a 10000 K and $\log(g) = 4.0$ atmosphere was simulated, resulting in an atmosphere thickness of $z = 4\times10^{10}$ cm, measured from $\tau_{Ross} =10^{-4}$ down to $\tau_{Ross} = 1$. We scaled this atmosphere to $T = 5000$ K using the pressure scale height, obtaining an atmosphere thickness $z' = 1.2 \times 10^8$ cm and a secondary's surface gravity $\log(g_s) \approx 4.6$ . The surface gravity value for the secondary in V3885 Sgr is consistent with an isolated M0V or M2V star \citep{All00}. Starting from the Roche lobe radius and the obtained atmosphere thickness, the average electron density is calculated (equation 2) as $N_{e} \approx 1.4 \times 10^{12}$ cm$^{-3}$. This value is an upper limit for the electron density, as an atmosphere expands as irradiated and, from equation 2, the electron density is inversely proportional to the square root of the emitting volume. Such a value is effectively within the optically thin region of our model atmosphere.

Using high time-resolution V-band data taken on a single run (Bruch 2006, private communication) we have calculated an average orbital light curve. The data comprise 12 orbital cycles and were binned into 40 phase bins using the orbital ephemeris proposed in this work. The average flux curve presents a maximum around phase 0.6 which is consistent with significant companion illumination effects in the optical continuum (fig. 10)\notetoeditor{Figure 10 must be a single column figure}. The possibility of the main line flickering feature being associated to the hot spot instead of the secondary can not be completely ruled out from the data at hand. However, it would imply in a large ($> 0.2$) spectroscopic phase shift. In this case the broadband light curve peak could not be associated with the secondary star. On the other hand, the expected absolute gas velocities at the hot spot are quite uncertain but possibly larger than those measured for the flickering feature at $V_x \approx$ 0.

\begin{figure}
\includegraphics[angle=270, width=85mm]{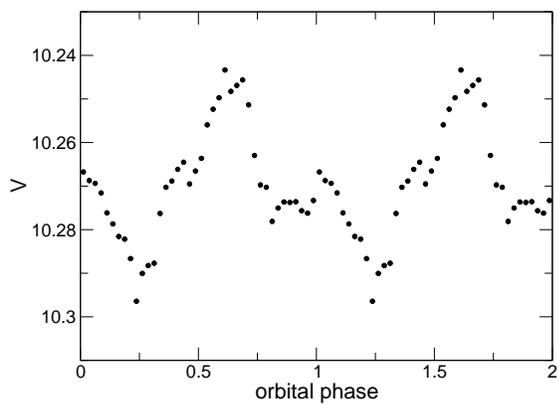}
\figcaption{V band orbital lightcurve of V3885 Sgr. Each point is an average of approximately 170 photometric datapoints comprising 12 orbital cycles.\label{fig10}}
\end{figure}

\placefigure{fig10}

From the arguments above, we suggest that the C$_1$ emission line flickering observed in V3885 Sgr could be due to a radiative variable component produced by recombination at the secondary's illuminated surface. The calculation described above provides only orders of magnitude, while a detailed model considering an irradiated atmosphere is still needed to ensure this conclusion. The observation of flickering coming directly from the inner accretion disc is not possible via flickering tomography because the information from higher velocities is diluted at the outer border of the tomograms. The time delay in recombination process could affect significantly only the flickering at high frequencies. As we used a integration time of about 3 minutes in our data, only the low frequency flickering component is mapped.

The possibility of the secondary be a flare star was also considered. Stellar flares are observed through all the main sequence, with the most intense flaring being dominant in the hotter stars (see \citet{Pet89} for a review). Stellar flares are observed mainly in cool stars, where the flares have smaller amplitude and are more frequent than the ones of the hot stars. The mechanism proposed to explain the flare production involves the magnetic field of the star. The dynamo mechanism is enhanced by the fast stellar rotation, for this reason it is expected that stars with higher rotation present enhanced flare activity \citep{Par55}. Cataclysmic Variables with their short orbital periods are the perfect scenario for this kind of activity. Stellar flares were observed in SDSS detached binary systems \citep{Sil06} and also in the magnetic CV AM Her during its low-state \citep{Kaf05}. Flickering and stellar flares present some differences: first, the flares appear as isolated outbursts while flickering is a "continuous" phenomenon and, second, the time-scales are distinct. AM Her has an orbital period of 3.8 h \citep{You81}, which is shorter than the V3885 Sgr orbital period. Therefore, the secondary in AM Her has higher rotation velocity to feed the dynamo and presents approximately one flare at each 5.4 d \citep{Kaf05}. As the less intense flares are more frequent, could the microflare phenomenon mimic the line flickering? \citet{Rob95} observed 32 flare events in 2 h observation of the microflaring dM8e star CN Leo. This represents a rate of 1 flare at every 3.7 min, which is considerably lower than the frequency of flickering events.

The energy associated with the individual events in stellar flares and flickering in CVs are also distinct. The typical luminosity of a V3885 Sgr line flickering event is calculated directly from the Doppler maps as $L \approx 4\times 10^{27}$ erg s$^{-1}$. Using the microflare time-scale of $\tau \approx 5s$ \citep{Rob99} and the energy range of $10^{24}-10^{25}$ erg \citep{Par88}, a luminosity range of $2 \times 10^{23}-2 \times 10^{24}$ can be inferred for the microflare events. The microflare energy range is therefore well below the flickering energy. The flare energy in red stars is below $10^{27}$ erg, this, combined with a typical time-scale of 10 s yields a luminosity below $10^{26}$ erg s$^{-1}$. This later value is also below the flickering flare energy found in the V3885 Sgr secondary star.

As the stellar flares are distinct from the flickering both in frequency and energy, this phenomenon does not seem to be the main cause of the C$_1$ emission line flickering observed in V3885 Sgr.

\section{Discussion}

The stellar masses are constrained within an interval of $0.3 < M_1 < 0.9$ M$_{\sun}$ and $0.25 < M_2 < 0.55$ M$_{\sun}$. The uncertainty in $K_1$ propagates to the derived mass ranges, so these ranges are just formal as well. This uncertainty does not include the fact that measured $K_1$ velocity in CVs may not reflect the white dwarf orbital velocity. \Citet{Har05} constrained the primary mass between 0.55 and 0.8 M$_{\sun}$, which is in agreement with our mass interval. Concerning the orbital inclination, we conclude that the possible values for the system are between 45$\degr$ e 75$\degr$. \citet{Cow77} used 60$\degr$ as a upper limit for the value of the orbital inclination in their mass diagram. \citet{Hau85} assumed that the orbital inclination is between 60$\degr$ and 80$\degr$. \citet{Har05} pointed that the orbital inclination must be higher than 65$\degr$. The uncertainty on this parameter remains high.

Regarding the morphology of Doppler reconstructions, an enhanced emission in the $(-V_x,+V_y)$ quadrant was observed both in H$\alpha$ and He\begin{small} I\end{small} tomograms. In both cases an enhanced emission in $(+V_x,-V_y)$ quadrant might be interpreted as being originated by the impact of the gas stream that have not reached the disc in the $(-V_x,+V_y)$ quadrant. Emission from the same region was searched by visually inspecting the Doppler Tomogram atlas by \citet{Kai94}. The presence of enhanced emission in the $(+V_x, -V_y)$ quadrant is seen in V1315 Aql (H$\beta$ and He\begin{small} I\end{small} 4471 lines), IP Peg (H$\beta$), LX Ser (He\begin{small} I\end{small} 4471), DW UMa (H$\beta$) and DX UMa (H$\beta$). From all these systems, only IP Peg is classified as a dwarf nova, while the other objects are classified as nova-like CVs. The orbital periods of all these systems are shorter than 5 hours. This effect may well be due to a selection effect as the catalogue has only 4 objects with orbital period longer than 5 hours and 14 objects with period shorter than 5 hours. High S/N tomography studies of other CVs are needed to investigate if there is any correlation between the orbital period or eruptive class and the presence of emission in the $(V_x,-V_y)$ quadrant. 

While the H$\beta$ and H$\gamma$ Doppler maps presented by \citet{Har05} have a ring signature, while our H$\alpha$ Doppler map is filled at low velocities. The obtained radial emissivity indexes for both H$\alpha$ and He\begin{small} I\end{small} lines are similar to the ones obtained for V841 Oph by \citet{Dia03}. Concerning the absolute intensities, V3885 Sgr is approximately 4 times fainter in the Balmer emission lines than V841 Oph.

V3885 Sgr is the second object targeted for a flickering tomography study, the first object was V442 Oph \citep{Dia01}. The flickering variance tomogram presents two regions of higher variability in the $(-V_x,+V_y)$ and $(V_x, -V_y)$ quadrants, which are not in the same location as the regions of higher emission. In the V442 Oph variance tomogram \citep{Dia01}, the region of most intense emission is in the $(-V_x,-V_y)$ and $(+V_x,-V_y)$ quadrants. The ratio between flickering and flux tomogram for V3885 Sgr and V442 Oph also presents a different behavior. For V442 Oph there is an region of low intensity near the center, while for V3885 Sgr the lowest intensity regions are not centered. In both cases it is verified that the flickering can vary with the position within the disc. The intrinsic distribution of the flickering as a function of the velocity also presents a distinct behavior for both systems. The flickering distribution along the V3885 Sgr disc showed to be non-homogeneous as observed in V442 Oph. A future study concerning methods for qualifying the significance of features in flickering maps is planned.

Comparing the flickering tomogram with simulations, we have associated one of the sources of emission line flickering with the secondary star. As far as we know, this is the first detection of flickering at the companion star in CVs. Previous eclipse mapping studies have located the continuum flickering sources restricted to the accretion disc in UU Aqr \citep{Bor06} and to the accretion disk and the gas stream in V2051 Oph \citep{Bap02b}. Using the flickering tomography method we are not able to observe the flickering originated in the inner parts of the accretion disc because the information at high velocities is spread at the outer radii of the Doppler map. In addition, the S/N ratio in the line wings is usually low. According to our simulations, the brightest line flickering source in V3885 Sgr is associated with the secondary star. One mechanism that may be proposed to explain this behavior is the flickering in the UV continuum from the accretion disc being reprocessed on the illuminated face of the secondary star. As the internal part of the accretion disc rotates faster than the disc's outer parts, the magnetic reconnection in the internal disc could be a mechanism to produce the high frequency flickering \citep{Gee92}. Flickering in the UV can be produced in the inner hot disc, resulting in an illumination effect in the optical region. The illumination of the atmosphere of the companion star results into both heating and ionization. Recombination in the optically thin atmosphere results in a low fraction of the illumination energy by emission lines \citep{Stri74}. \citet{Har05} presented evidences of illumination of the secondary star in V3885 Sgr, as a narrow emission component superposed to the lines and emission from the illuminated face of the secondary star in the H$\beta$ and H$\gamma$ Doppler maps. We present photometric data that supports the scenery of a illuminated companion star. In our opinion, it would be much more difficult to argue in favor of the hot spot as responsible for the C1 feature. The absolute velocity for the hot spot would be too low and it would require a large phase shift $(> 0.2)$.

\section{Conclusions}

We present the analysis of an extensive spectroscopic observation of V3885 Sgr. A long term ephemeris was determined for V3885 Sgr, with orbital period $P_{orb}$ = 0.20716071(22) d, and spectroscopic conjunction, $E_0$ = 51942.08371(46) MHJD. The radial velocity semiamplitudes and the mass ranges derived by \citet{Har05} are confirmed by our measurements.

Using flickering tomography, it was verified that the flickering intensity can vary within the disc. From the V3885 Sgr flickering tomograms emissivity profiles, one can see that the flickering is most intense at low velocity regions. From the flickering tomography studies of V3885 Sgr and V442 Oph, one can conclude that the line flickering in CVs may have a non-homogeneous distribution along the disc. More additional flickering tomography studies from other systems are needed to ensure this conclusion. Comparing synthetic flickering tomograms with observed ones, we concluded that the most intense line flickering source could not be originated in a Keplerian accretion disc defined by the V3885 Sgr mass interval and orbital period. The main flickering source may be associated with the secondary star, the origin of such feature may be the flickering in the UV continuum produced in the accretion disc being reprocessed at the illuminated face of the companion star.

\acknowledgments

This work is based on data obtained at LNA/CNPq and Cerro Tololo observatories. F.M.A.R is grateful from support from FAPESP fellowship 01/07078-8. MD acknowledges the support by CNPq under grant \#301029. We wish to thank Ruth Gruenwald for her carefull reading of the manuscript.

\clearpage

\clearpage

\clearpage

\clearpage

\clearpage

\clearpage

\clearpage

\clearpage

\begin{table*}
\begin{center}
\caption{Journal of Observations\label{tbl-1}}
\begin{tabular}{rccccc}
\tableline\tableline
observation date & telescope & exp. time (s) & n. of spectra & n. of cycles \\
\tableline
1999 Sep 01 & 1.60-m LNA  & 45  & 280	& 1.17 \\
1999 Sep 02 & 1.60-m LNA  & 45  & 390	& 1.55 \\
2000 Jul 07 & 1.60-m LNA  & 30  & 148	& 0.46 \\
2000 Jul 08 & 1.60-m LNA  & 30  & 72	& 0.28 \\
2000 Jul 09 & 1.60-m LNA  & 30  & 134	& 0.54 \\
2001 Mar 21 & 1.60-m LNA  & 45  & 112	& 0.45 \\
2001 Mar 22 & 1.60-m LNA  & 45  & 62	& 0.28 \\
2001 Sep 10 & 1.60-m LNA  & 90  & 128	& 1.13 \\
2001 Sep 11 & 1.60-m LNA  & 90  & 30	& 0.31 \\
2001 Sep 18 & 1.60-m LNA  & 100 & 117	& 1.06 \\
2001 Sep 19 & 1.60-m LNA  & 100 & 89	& 0.88 \\
2002 Jun 16 & 1.5-m CTIO  & 180 & 25	& 0.26 \\
2002 Jun 17 & 1.5-m CTIO  & 180 & 25	& 0.26 \\
2002 Jun 18 & 1.5-m CTIO  & 180 & 32	& 0.38 \\
2002 Jun 19 & 1.5-m CTIO  & 180 & 30	& 0.35 \\
2002 Jun 20 & 1.5-m CTIO  & 180 & 37 	& 0.39 \\
2002 Jun 21 & 1.5-m CTIO  & 180 & 36	& 0.38 \\
2002 Jun 22 & 1.5-m CTIO  & 180 & 29	& 0.47 \\
2002 Jun 23 & 1.5-m CTIO  & 180 & 42	& 0.46 \\
\tableline
\end{tabular}
\end{center}
\end{table*}

\end{document}